\documentclass[aps,prb,showpacs,twocolumn,superscriptaddress]{revtex4}
\usepackage{amssymb}
\usepackage{graphicx}% Include figure files
\usepackage{dcolumn}% Align table columns on decimal point
\usepackage{bm}% bold math
\usepackage{color}
\begin{document}

\title{Fast preparation of single hole spin in InAs/GaAs quantum dot in Voigt geometry magnetic field}

\author{T.~M.~Godden}
\affiliation{Department of Physics
and Astronomy, University of Sheffield, Sheffield, S3 7RH, United
Kingdom}

\author{J.~H.~Quilter}
\affiliation{Department of Physics
and Astronomy, University of Sheffield, Sheffield, S3 7RH, United
Kingdom}

\author{A.~J.~Ramsay}
\email{a.j.ramsay@shef.ac.uk}
\affiliation{Department of Physics
and Astronomy, University of Sheffield, Sheffield, S3 7RH, United
Kingdom}

\author{Yanwen Wu}
\affiliation{Cavendish Laboratory, University of Cambridge, Cambridge, CB3 OHE, United
Kingdom}
\affiliation{Department of Physics,
The University of Texas at Austin,
Austin, Texas 78712, USA}

\author{P.~Brereton}
\affiliation{Cavendish Laboratory, University of Cambridge, Cambridge, CB3 OHE, United
Kingdom}

\author{I.~J.~Luxmoore}
\affiliation{Department of Physics and Astronomy, University of
Sheffield, Sheffield, S3 7RH, United Kingdom}

\author{J.~Puebla}
\affiliation{Department of Physics and Astronomy, University of
Sheffield, Sheffield, S3 7RH, United Kingdom}

\author{A.~M.~Fox}
\affiliation{Department of Physics and Astronomy, University of
Sheffield, Sheffield, S3 7RH, United Kingdom}

\author{M.~S.~Skolnick}
\affiliation{Department of Physics and Astronomy, University of
Sheffield, Sheffield, S3 7RH, United Kingdom}

\date{\today}% It is always \today, today,

\begin{abstract}
The preparation of a coherent heavy-hole spin via ionization of a spin-polarized electron-hole pair in an InAs/GaAs quantum dot in a Voigt geometry magnetic field is experimentally investigated. For a dot with a typical bright-exciton fine-structure splitting of $17~\mathrm{\mu eV}$ , the fidelity of the spin preparation is limited to 0.75, with optimum preparation occurring when the effective fine-structure of the bright-exciton matches the in-plane hole Zeeman energy. In principle, higher fidelities can be achieved by minimizing the bright-exciton fine-structure splitting.
\end{abstract}
\maketitle

\section{Introduction}

The spin of a carrier confined to a self-assembled semiconductor quantum dot is a promising qubit with lifetimes  \cite{Kroutvar_nat,Heiss_prb} and intrinsic coherence times in the ms and $\mathrm{\mu s}$ \cite{Greilich_sci,Press_nphoton,Brunner_sci} regime respectively.
An important first step of any quantum information protocol or coherent control experiment is to prepare the spin in a well-defined initial state. Broadly speaking, there are three classes of optical spin initialization technique that have so far been experimentally realized, which we term: spin pumping \cite{Greilich_sci,Greilich_prl,Mikkelsen_nphys,Berezovsky_sci,Eble_prl,Fras_apl}, spin distillation \cite{Atature_sci,Gerardot_nat,Xu_prl} and exciton ionization \cite{Kroutvar_nat,Heiss_prb,Ramsay_prl1,Godden_apl,Young_njp}. For a review of spin preparation techniques, see ref. \cite{Ramsay_sst}.

To prepare a spin using exciton ionization, a spin-polarized electron-hole pair is optically generated in a dot with a strong asymmetry in the electron/hole tunneling rates. Under an applied electric-field one of the carriers, for example the electron,  tunnels from the dot to leave a spin-polarized hole. This approach was first demonstrated on ensembles of dots using a voltage-gated photoluminescence detection technique \cite{Kroutvar_nat,Heiss_prb}, and later extended to a single electron \cite{Young_njp,Heiss_prb2}. Using resonant \cite{Ramsay_prl1}, rather than quasi-resonant excitation, and a Faraday geometry magnetic field, high-fidelity ($>99\%$), high-speed ($1/e$ time $30~\mathrm{ps}$), spin preparation has recently been achieved \cite{Godden_apl} using a photocurrent detection technique.

Recently, we demonstrated the coherent optical control of a single hole spin using photocurrent detection \cite{Godden_prl}. Similar reports were made around the same time by independent groups using different methods \cite{DeGreve_nphys,Greilich_nphoton}. In our experiments, Larmor precession of the hole spin about an in-plane magnetic field was observed, signifying the preparation of a coherent hole-spin using exciton ionization. %In %these experiments, the 100-ps timescale of the electron tunneling rate, compared to the 180-ps period of the exciton spin precession should be %sufficient to fully entangle the electron and hole spins. Therefore one might expect this to  lead to a mixed hole-spin state when the electron %tunnels from the dot.

Here we present an investigation into the preparation of a coherent hole-spin that Larmor precesses in a Voigt geometry magnetic field following ionization of a spin-polarized electron-hole pair prepared by resonant excitation. The contrast of the Larmor precession is found to increase with both applied magnetic and electric fields. A model is developed to explain the observed behavior. According to the model, a circularly-polarized laser pulse prepares a spin-up bright-exciton. This is a superposition of the bright-exciton energy eigenstates separated by an effective fine-structure splitting that is a function of applied magnetic field. In the presence of the electron, the hole spin acquires a rotating and counter-rotating phase-factor. When the effective fine-structure splitting and the in-plane hole Zeeman energy are close to resonance, the rotating component is synchronized with the Larmor precession of the hole spin and is unaffected by the tunneling of the electron, whereas the phase of the counter-rotating hole-spin component is washed out by the uncertainty in the timing of the electron tunneling event. This leads to a Larmor precession of maximum contrast 0.5, or fidelity of 0.75. In principle, a route to higher fidelity spin preparation is to use a dot with near zero fine-structure splitting \cite{Young_prb}, and a small magnetic field to minimize the desynchronization of the rotating and counter-rotating components of the hole spin in the presence of the electron.

The experiments presented here were performed on a single InAs/GaAs quantum dot embedded in the intrinsic region of an ni-Schottky diode. Details of the device structure can be found in ref. \cite{Boyle_prb}. A photocurrent detection technique is used \cite{Zrenner_nat,Zecherle_prb,Takaghi_oe,Wu_prl}. A background signal proportional to the incident power \cite{Stufler_prb} has been subtracted for all data presented. Full details of the optical setup can be found in the appendix.

\section{Principle of operation}

\begin{figure}
\begin{center}
\includegraphics[scale=0.4]{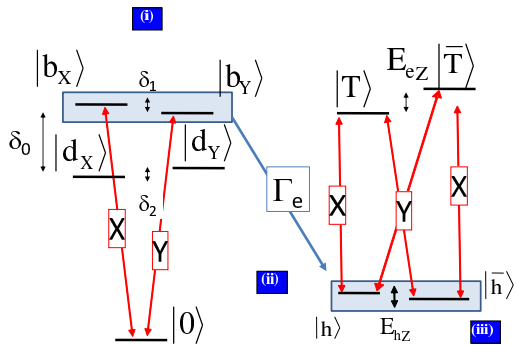}
\end{center}
\caption{ Energy-level diagram of neutral exciton and positive trion sub-systems in a Voigt geometry magnetic field. The neutral exciton has two bright $m_J=\pm 1$ states $\vert b\rangle$ and two dark $m_J=\pm 2$ states $\vert d\rangle$ that are mixed by the magnetic field. The hole (trion) spin energy eigen-states are aligned parallel $\vert h\rangle ~(\vert T\rangle)$ and anti-parallel $\vert \bar{h}\rangle ~(\vert \bar{T}\rangle) $ to the magnetic field.
}\label{fig:fig1a}
\end{figure}

Figure \ref{fig:fig1a} presents an energy-level diagram of the neutral exciton and hole-trion sub-systems in a Voigt geometry magnetic field. At zero magnetic field, the bright $m_J=\pm 1$ and dark neutral excitons $m_J=\pm 2$ are separated by the exchange energy $\delta_0$. The bright-exciton states, $\vert b_{x,y}\rangle=(\mid \downarrow\Uparrow\rangle \pm \mid \uparrow\Downarrow\rangle)/\sqrt{2}$, are separated by the anisotropic exchange energy $\delta_1$, and the dark-exciton states ($\vert d_{x,y}\rangle=(\mid \uparrow\Uparrow\rangle \pm \mid \downarrow\Downarrow\rangle)/\sqrt{2}$) are separated by $\delta_2$. The bright-exciton states are optically active, and are labeled by their linearly polarized selection rules. Application of a Voigt geometry magnetic field couples bright and dark excitons of the same symmetry. The bright and dark excitons are mixed, and at high magnetic field the dark exciton becomes optically accessible. The energy splitting between the  bright-excitons, the effective fine structure splitting $E_{bb}\approx \delta_1+KB^2$, is a function of magnetic field \cite{Stevenson_prb}. For the dot investigated here, $\delta_1$ and $K$ have the same sign, and $KB^2\ll \delta_1$.

When the electron tunnels from the dot, the dot is positively charged, and the energy-levels of the hole-trion system are also presented in fig. \ref{fig:fig1a}. The hole-spin states $\vert h,\bar{h}\rangle= (\mid \Uparrow\rangle \pm \mid \Downarrow\rangle)/\sqrt{2}$ are aligned parallel and anti-parallel to the applied magnetic field and are separated by the in-plane hole Zeeman energy $E_{hZ}=-g_{hx}\mu_B B$. Likewise, the positive trion states $\vert T,\bar{T}\rangle= (\mid \Uparrow\Downarrow\uparrow\rangle \pm \mid \Uparrow\Downarrow\downarrow\rangle)/\sqrt{2}$ are also aligned parallel and anti-parallel to the magnetic field and are separated by the in-plane electron Zeeman energy $E_{eZ}=-g_{ex}\mu_BB$. The optical selection rules are linearly polarized, with x-polarized light preserving the symmetry of the state, and y-polarized light flipping the symmetry.

\begin{figure}
\begin{center}
\includegraphics[scale=0.35]{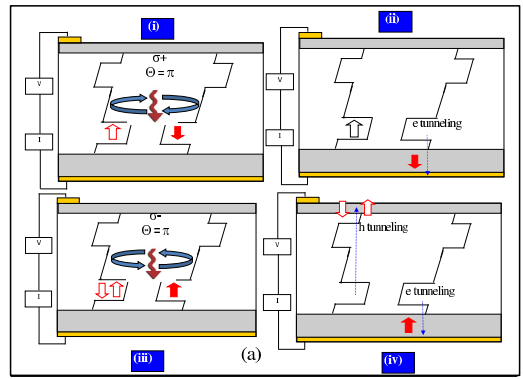}
\end{center}
\caption{(a) Principle of operation. {\it Preparation} (i) Resonant excitation with $\sigma_+$-polarized $\pi$-pulse prepares a spin-polarized electron-hole pair in the dot. (ii) The electron tunnels to leave a hole with a net spin. {\it Detection} (iii) Resonant excitation of the hole-trion transition with a circularly polarized $\pi$-pulse creates a trion conditional on the spin state of the hole. (iv) Eventually, all carriers tunnel from the  dot leading to a change in photocurrent proportional to the occupation of the target hole spin state when the detection pulse arrived.
}\label{fig:fig1b}
\end{figure}

Figure \ref{fig:fig1b} presents a schematic of the principle of operation. An electric-field is applied such that the electron tunneling rate $(1/(100~\mathrm{ps}))$ is fast compared to the hole tunneling rate $(1/(4~\mathrm{ns}))$. A magnetic field is applied in the plane. The dot is excited by two circularly-polarized laser pulses of pulse-area $\pi$, termed preparation and detection. They have a Gaussian shape and intensity FWHM of 0.2 meV corresponding to 14~ps FWHM in the time-domain. The preparation pulse excites the bright-exciton transition on-resonance, creating a spin-up electron-hole pair. This is a superposition of the bright-exciton states which causes the exciton spin to precess at the effective fine-structure splitting, which is approximately $17~\mathrm{\mu eV}$ at low magnetic field. For reasons that will be discussed later, when the electron tunnels it leaves a hole with a net spin precession synchronized with the preparation pulse. The hole spin will then Larmor precess about the external magnetic field, at the in-plane hole Zeeman energy. To detect the hole spin, the detection pulse excites the hole-trion transition on-resonance. Due to Pauli exclusion, circular polarization selects a hole-spin to map to the trion state. In time, the additional photo-generated carriers tunnel from the dot giving rise to a change in photocurrent proportional to the occupation of the polarization selected hole-spin state at the instant when the detection pulse arrived.

\section{Results}

\begin{figure}
\begin{center}
\includegraphics[scale=2.5]{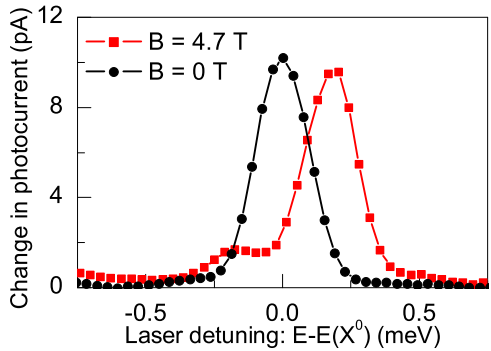}
\end{center}
\caption{(a) Photocurrent spectra of neutral exciton transition with and without application of a 4.7-T Voigt geometry magnetic field.
At high magnetic field, an additional peak is observed due to mixing of the bright and dark excitons that activates the  dark-exciton state.
(b) Diamagnetic shift of neutral exciton transition with in-plane magnetic field $B$.
}\label{fig:fig2}
\end{figure}

First the wavelength to be used for the preparation pulse is determined. This is achieved by exciting the dot with the preparation pulse only, and measuring a photocurrent spectrum of the neutral exciton transition by scanning the frequency of the laser. Example spectra are presented in fig. \ref{fig:fig2}(a). At zero magnetic field, a single bright-exciton peak is observed. With increasing magnetic field, the bright and dark excitons are mixed, and at high magnetic field a weaker  dark-exciton peak is also observed \cite{Bayer_prb}. From the energy splitting of the bright and dark exciton, we estimate the bright-dark exciton exchange energy of $\delta_0=315\pm 16~\mathrm{\mu eV}$.

Figure \ref{fig:fig2}(b) presents the diamagnetic shift of the energy of the bright neutral-exciton transition, fitted to $E(X^0)=E_0+\chi B^2$, where $\chi=5.3\pm 0.3 ~\mathrm{\mu eV.T^{-2}}$. From the diamagnetic-shift $\chi$ it is possible to estimate the radius of the exciton wavefunction, using $r_0=\sqrt{\frac{8\mu \chi}{e^2}}$, where $\mu$ is the exciton effective mass in GaAs and assuming a spherical Gaussian wavefunction $\psi(r)\propto e^{-r^2/2r_0^2}$. \cite{Coey_text} A value of $r_0=3.5\pm 0.1 ~\mathrm{nm}$ is deduced, consistent with the value deduced for similar dots from the intensity damping of exciton Rabi rotations \cite{Ramsay_prl2010}.

\begin{figure}
\begin{center}
\includegraphics[scale=2]{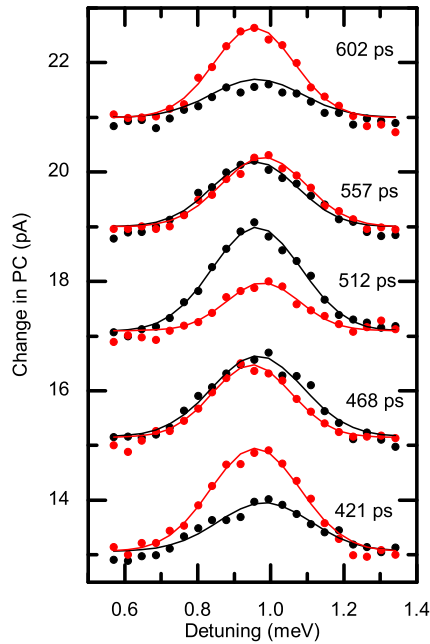}
\end{center}
\caption{Larmor precession of heavy-hole in 4.7-T magnetic field. Two-color photocurrent spectra are measured by tuning the detection pulse through the hole-trion transition for co (red) and cross-circularly (black) polarized excitation. The co/cross-circular trion peaks oscillate in anti-phase and are displayed for various values of the inter-pulse time-delay over one 180-ps period of the Larmor precession.
}\label{fig:fig3}
\end{figure}

To measure the Larmor precession of the heavy-hole, the dot is excited by a preparation and detection pulse. The preparation pulse is on-resonance with the bright-exciton peak to prepare the hole-spin. A two-color photocurrent spectrum is then measured by scanning the frequency of the detection pulse through the hole-trion transition, resulting in an additional peak about 1-meV higher in energy than the bright exciton that is not observed in the one pulse experiments. Figure \ref{fig:fig3} presents a series of two-color photocurrent spectra for co (red) and cross-circular (black) excitation for various inter-pulse time-delays, at a magnetic field of 4.7 T, and a reverse bias of 1.0~V. For a $\sigma_+$ preparation pulse, the amplitudes of the co/cross circular are proportional to the occupations of the hole spin down/up states respectively. These peaks oscillate in anti-phase. The position of the hole-trion peaks also oscillates with time-delay. We speculate that this oscillation may be due to optical pumping of the nuclear spins, where the pumping rate is modulated by the inter-pulse time-delay \cite{Ladd_prl,Makhonin_apl}.

\begin{figure}
\begin{center}
\includegraphics[scale=1.3]{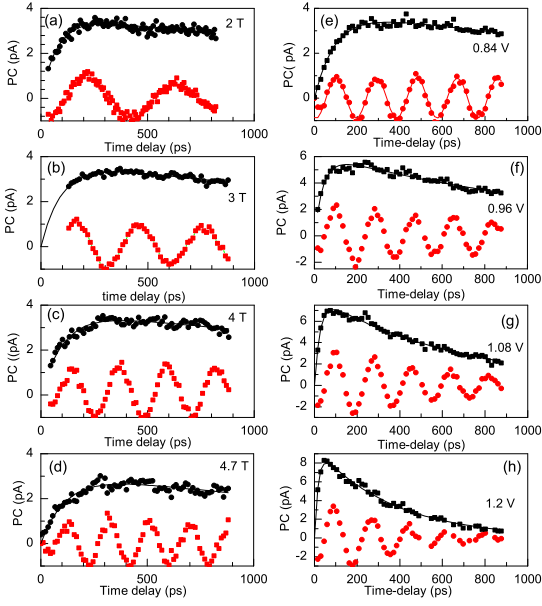}
\end{center}
\caption{(a-d) Larmor precession of hole-spin vs magnetic field at $V_g=0.80~\mathrm{V}$. (e-h) Larmor precession of hole-spin vs gate voltage at $B=4.7~\mathrm{T}$. (black/red) Sum/difference of amplitudes of trion peak for co/cross circular excitation conditions. (black) The sum is proportional to the total hole population. (red) The difference is proportional to the hole spin-z component.
}\label{fig:fig4}
\end{figure}

\begin{figure}
\begin{center}
\includegraphics[scale=2]{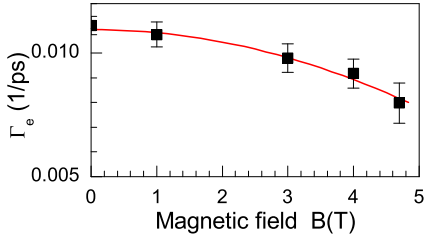}
\end{center}
\caption{Electron tunneling rate vs applied magnetic field. The tunneling rate decreases due to an effective increase in the barrier thickness due to the Lorentz force experienced by carriers tunneling from the dot.}
\label{fig:fig5}
\end{figure}

Figure \ref{fig:fig4}(a-d) presents a series of Larmor precession measurements for various applied magnetic fields. In these measurements a series of two-color photocurrent spectra are measured as in fig. \ref{fig:fig3}, and the amplitudes of the co/cross circular excitation peaks are extracted from fits to Gaussian functions. The black traces present the sum of the co/cross circular amplitudes, which is proportional to the total hole population. At short time-delays, the signal rises exponentially as the electron tunnels from the dot and then decays exponentially as the hole tunnels from the dot, ie. the sum signal $n_h\propto (e^{-\Gamma_h \tau}-e^{-\Gamma_e\tau})$, where $\Gamma_e,\Gamma_h$ are the electron and hole tunneling rates respectively, and $\tau$ is the inter-pulse time-delay.

As the magnetic field is increased, the tunneling rates decrease since the carriers experience a Lorentz force when tunneling increasing the effective barrier width. The magnetic field dependence of the electron tunneling rate is plotted in fig. \ref{fig:fig5}, and fitted to a WKB expression \cite{Patane_prb}:
\begin{equation}
\frac{\Gamma_e(B)}{\Gamma_e(0)}=1-\frac{16U^{5/2}_I (eB)^2}{15\hbar(2m^*)^{1/2}(eF)^3}+O(B^4)
\end{equation}
where $U_I$ is the barrier height, and $m^*=0.063m_0$ is the effective mass of the electron. $F=(V_g+V_{bi})/L$ is the applied electric-field, where $V_{bi}=0.76~\mathrm{V}$, is the built in voltage, approximated as half the GaAs bandgap, $L=230~\mathrm{nm}$ is the separation of the contacts. A value of $U_I=165\pm 10 ~\mathrm{meV}$ is deduced from the fit, which is consistent with the value of $171\pm 10~\mathrm{meV}$ deduced from the voltage dependence of the electron tunneling rate fitted using the WKB expression given  in ref. \cite{Fry_apl}.

\begin{figure}
\begin{center}
\includegraphics[scale=2]{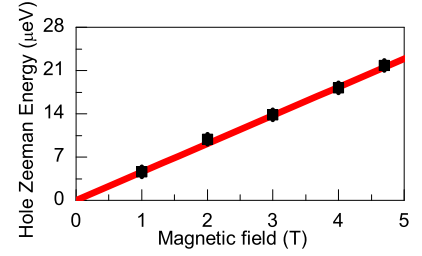}
\end{center}
\caption{Magnetic field dependence of Larmor precession frequency of hole spin at $V_g= 0.84~\mathrm{V}$. An in-plane hole g-factor of $\vert g_{hx}\vert =0.079\pm 0.004$ is deduced.
}\label{fig:fig6b}
\end{figure}

The red-traces of fig. \ref{fig:fig4} present the difference between the co and cross circular amplitudes, which is proportional to the spin-z component of the hole. As presented in fig. \ref{fig:fig6b}, the frequency of the oscillation is proportional to the magnetic-field, confirming that the oscillation is due to Larmor precession of the hole-spin with an in-plane g-factor of $ \vert g_{hx}\vert =0.079\pm 0.004 $. An in-plane g-factor indicates some mixing of the light and heavy-hole. Recent work has shown that the in-plane hole g-factor can be highly anisotropic \cite{Schwan_apl}; larger values may thus be accessible in our sample for different in-plane field direction.

\begin{figure}
\begin{center}
\includegraphics[scale=2]{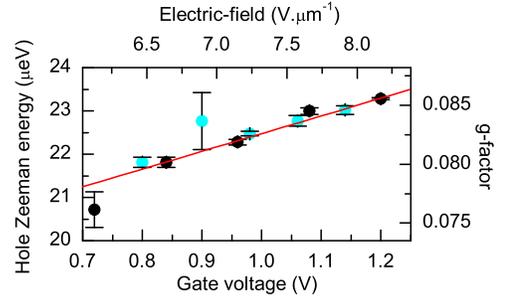}
\end{center}
\caption{Electric-field dependence of in-plane hole g-factor. The hole Zeeman energy is fitted to $E_{hZ}=(18.4\pm 0.2) + (4.1\pm 0.2) V_g ~\mathrm{\mu eV}$.
}\label{fig:fig6}
\end{figure}

Figure \ref{fig:fig4}(e-h) presents a series of Larmor precession measurements of the hole spin for various gate voltages at $B=4.7~\mathrm{T}$. As the reverse bias is increased, the carrier tunneling rates increase leading to faster exponential rise and fall in the signal. The maximum photocurrent signal observed for the hole-trion transition also increases, since the photocurrent signal is limited by the need for all carriers to tunnel before the arrival of the next pulse sequence \cite{Kolodka_prb}. The in-plane hole Zeeman energy increases linearly with gate voltage by up to 9\%, as shown in fig. \ref{fig:fig6}. This is attributed to an in-plane hole g-factor that depends on vertically applied electric-field $F$ as $\vert g_{hx}(F)\vert =(0.051 \pm 0.006) + (0.0035 \pm 0.0002 ~\mathrm{V^{-1}.\mu m}) F$. Similar observations have been made in other Larmor precession measurements, where the extrinsic dephasing of the hole-spin was attributed to charge induced fluctuations in the g-factor \cite{Greilich_nphoton,DeGreve_nphys}. In ref. \cite{Klotz_apl,Jovanov_prb}, a large 250\% decrease in the out-of-plane g-factor of the exciton was observed with vertically applied electric-field.  This was attributed to changes in the hole g-factor due to changes in the overlap of the hole wavefunction with Indium rich/poor regions of the dot. A similar effect could probably explain the change in the in-plane hole g-factor, observed here. An 8\% change in  the out-of-plane g-factor of the exciton has also been observed in a lateral electric-field \cite{Nakaoka_prb}. Another possibility is that the observed g-factor is influenced by the gate-voltage dependent optical pumping of the nuclear spin bath \cite{Makhonin_apl}.

\begin{figure}
\begin{center}
\includegraphics[scale=2.3]{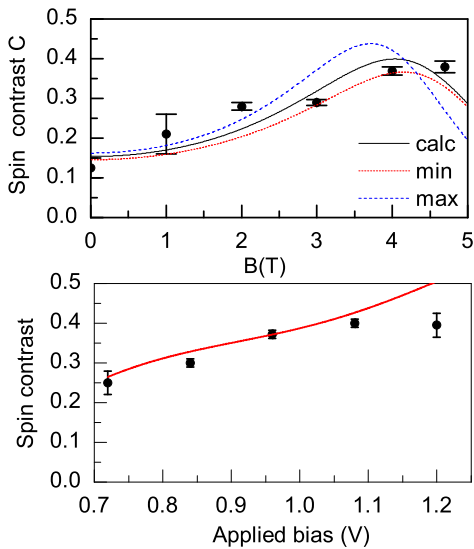}
\end{center}
\caption{Magnetic and electric field dependence of maximum contrast of Larmor precession. Lines give calculation of contrast using measured parameters.
}\label{fig:fig7}
\end{figure}

Figure \ref{fig:fig7} presents the contrast of the Larmor precession as a function of magnetic and electric fields.
The contrast of the Larmor precession is defined as $C=\lim_{\Gamma_e\tau \gg 1}\frac{PC_{+-}-PC_{++}}{PC_{+-}+PC_{++}}\vert_{max}$, where $PC_{+-}$ and $PC_{++}$ are the photocurrents measured for cross and co-circular excitation respectively. The contrast is a measure of the purity of the hole spin  prepared through the ionization of the exciton, with a value of one being ideal. The contrast improves with both increasing magnetic and electric-fields. A faster electron tunneling rate reduces the time for the exciton spin to evolve, hence the improvement with electric-field.
The increase in the contrast with magnetic field is at first surprising, since the effective fine-structure-splitting $E_{bb}$, the energy splitting between the bright-exciton states increases with magnetic field for this dot. Therefore one might expect the contrast to decrease with magnetic field as a result of faster precession of the exciton spin. In the next section, a more careful analysis will reveal why the contrast increases.
The lines are calculations using the model and independently measured dot parameters. The lines are calculated for parameters values at the extremes of the error bars, to present the full range of possible values for the contrast. The calculations will be discussed below.

\section{Model}

We adapt the model of ref. \cite{Godden_apl} to the case of an in-plane magnetic field.
In a magnetic-field $B$ applied along the x-axis, where the growth and optical-axis are aligned along the z-direction the Hamiltonian for the neutral exciton, as sketched in the energy-level diagram of fig. \ref{fig:fig1a} is \cite{Bayer_prb}:

\begin{eqnarray}
H^{0}= \frac{1}{2}\sum_{\alpha=x,y}[(\delta_0\pm\delta_1)\vert b_{\alpha}\rangle+(-\delta_0\pm \delta_2)\vert d_{\alpha}\rangle \nonumber \\ +\mu_BB(g_{ex}\mp g_{hx})\vert b_{\alpha}\rangle\langle d_{\alpha}\vert +h.c.]
\end{eqnarray}
%where $\vert B_{x,y}\rangle=[\vert \downarrow\Uparrow\rangle \pm \vert\uparrow\Downarrow\rangle]/\sqrt{2}$ are the bright ($m_J=\pm 1$ neutral %exciton states and $\vert D_{x,y}\rangle=[\vert \uparrow\Uparrow\rangle \pm \vert\downarrow\Downarrow\rangle]/\sqrt{2}$ are the dark ($m_J=\pm 2$) %neutral exciton states, composed of an electron $m_s=\pm 1/2$ and a heavy-hole $m_J=\pm 3/2$. $\delta_0$ are the exchange and $\delta_1,\delta_2$ %anisotropic exchange energies, which give rise to energy splittings between the bright/dark, bright-bright and dark-dark states respectively. %$g_{ex},g_{hx}$ are the in-plane g-factors of the electron and heavy-hole respectively. The in-plane magnetic field couples the bright and dark %excitons while conserving the symmetry of the wavefunction.

The energy-eigenstates of $H^0$ are:
\begin{eqnarray}
\vert \psi_{\alpha}(b)\rangle = \cos{\theta_{\alpha}}\vert b_{\alpha}\rangle +\sin{\theta_{\alpha}}\vert d_{\alpha}\rangle \\
\vert \psi_{\alpha}(d)\rangle = -\sin{\theta_{\alpha}}\vert b_{\alpha}\rangle +\cos{\theta_{\alpha}}\vert d_{\alpha}\rangle
\end{eqnarray}
\noindent where $\theta_{\alpha}$ is the mixing angle of the bright and dark excitons. The  bright exciton states are separated in energy by the effective fine-structure splitting $E_{bb}$, a key parameter for determining the purity of the spin preparation \cite{Stevenson_prb}:
\begin{eqnarray}
E_{bb}=\hbar\omega_{bb}\approx\delta_1 +KB^2
\end{eqnarray}

At time $t=0$, a $\sigma_+$-polarized laser pulse resonantly excites the bright exciton states to create a superposition state such that:
\begin{eqnarray}
\vert \psi(t)\rangle = \cos{\phi}e^{i\omega_{bb}t/2}\vert \psi_x(b)\rangle+\sin{\phi}e^{-i\omega_{bb}t/2}\vert\psi_y(b)\rangle \\
\tan{\phi}=\frac{\cos{\theta_x}}{\cos{\theta_y}}
\end{eqnarray}
where $\phi$ is set to minimize the initial occupation of the $\mid \uparrow\Downarrow\rangle$ state.

Hence the occupation of the hole-spin up/down exciton states $s_{\Uparrow,\Downarrow}$ evolve as:
\begin{eqnarray}
s_{\Uparrow}-s_{\Downarrow}=\sin{2\phi}\cos{(\theta_x+\theta_y)}\cos{(\omega_{bb}t)}e^{-\Gamma t} \\
s_{\Uparrow}+s_{\Downarrow}=e^{-\Gamma t}
\end{eqnarray}
\noindent where a decay rate $\Gamma=\Gamma_e+\Gamma_r$ due to the phenomenological electron tunneling rate $\Gamma_e$ and the radiative recombination rate $\Gamma_r$ has been added.

Next, rate equations for the occupations of the hole spin states $h_{\Uparrow,\Downarrow}$ are constructed, where $\Gamma_h$ is the hole tunneling rate and $\hbar\omega_{hZ}=-g_{hx}\mu_BB$ is the in-plane hole Zeeman energy:
\begin{eqnarray}
\dot{h}_{\Uparrow}-\dot{h}_{\Downarrow}=-(\Gamma_h+i\omega_{hZ})(h_{\Uparrow}-h_{\Downarrow})+\Gamma_e(s_{\Uparrow}-s_{\Downarrow}) \label{eq:hspin} \\
\dot{h}_{\Uparrow}+\dot{h}_{\Downarrow}=-\Gamma_h(h_{\Uparrow}+h_{\Downarrow})+\Gamma_e(s_{\Uparrow}+s_{\Downarrow})
\end{eqnarray}

In Eq. \ref{eq:hspin}, in the absence of the exciton spin, the hole-spin rotates with a phase-factor $e^{-i\omega_{hZ}t}$ about the external magnetic field. The hole-spin is driven by an exciton spin that has rotating and counter-rotating phase-factors $e^{\pm i\omega_{bb}t}$, since the sense of rotation of the hole spin in the presence of the electron is undefined. For $\omega_{bb} \sim \Gamma_e$, efficient transfer of the spin from one rotation component of exciton to the hole occurs when the hole Zeeman $\omega_{hZ}$ and effective fine structure splitting $\omega_{bb}$ are frequency matched. A possible physical interpretation is that when the electron tunnels from the dot, the sense of rotation experienced by the hole-spin in the presence of the electron is measured in a basis where the outcomes are aligned or anti-aligned with respect to the external magnetic field.
The equations are now integrated, and we take the real-part. The contrast is defined such that:
\begin{eqnarray}
C=\lim_{\Gamma_e t\gg 1\gg \Gamma_h t} \frac{h_{\Uparrow}-h_{\Downarrow}}{h_{\Uparrow}+h_{\Downarrow}} \\
C=\frac{1}{2}\sin{2\phi}\cos{(2\theta)}[f(\omega_{hZ}-\omega_{bb})+f(\omega_{hZ}+\omega_{bb})] \label{eq:C}\\
f(a)=\frac{(\Gamma-\Gamma_h)^2}{(\Gamma-\Gamma_h)^2+a^2}
\end{eqnarray}
\noindent where $2\theta=\theta_x+\theta_y$, and for most cases of interest $\theta_x\approx\theta_y$, and therefore $\phi\approx \pi/4$.

Two factors determine the spin contrast. A competition between the electron tunneling rate and the frequency mismatch between the in-plane hole Zeeman energy and the effective fine-structure splitting $E_{bb}$; and a factor that describes the mixing of the bright and dark-excitons. In eq. \ref{eq:C}, the curve $f(a)$ is similar to a Hanle depolarization curve and has similar dynamical origins. In a typical Hanle experiment, a quantum dot is excited with circularly polarized light, and the degree of circular polarization of the photoluminescence is measured as a function of in-plane magnetic field, see for example refs. \cite{Krebs_prl,Oulton_prl}. The classical Hanle depolarization curve is a Lorentzian, resulting from the competition between radiative recombination of the trion and the depolarization of the trion spin due to the Larmor precession. In contrast to a Hanle depolarization curve, in principle the resonance here occurs at a non-zero magnetic field,  since we measure the efficiency of transferring spin from an exciton to a hole precessing at different frequencies, where the most efficient transfer occurs when the precessions of the exciton and hole spins are synchronized.

The overall efficiency of the spin initialization is limited by the probability of creating a hole, $p_h=\frac{\Gamma_e}{\Gamma-\Gamma_h}$, which in turn, is limited by radiative recombination. For a typical dot with radiative lifetimes ranging from 500-1000~ps \cite{Malik_prb}, and electron tunneling time of 30-100 ps, a hole is created with an estimated probability of 83-97\%.

\section{Comparison of spin preparation model to data}

The lines presented in fig. \ref{fig:fig7} are calculations of the contrast using eq. \ref{eq:C}. The traces are calculated to give maximum and minimum values of the contrast using measured dot parameters. According to Eq. \ref{eq:C}, the contrast depends on the hole Zeeman energy $E_{hZ}$, the effective fine-structure splitting of the bright exciton $E_{bb}(B)$, the electron tunneling rate $\Gamma_e$ and the mixing of the bright and dark neutral excitons $\cos{(2\theta)}$. The in-plane hole g-factor is deduced from the period of the Larmor precession of the hole spin. The effective fine-structure splitting was measured as function of magnetic field by measuring the precession of the exciton spin. Time resolved measurements of the exciton spin precession were achieved by measuring photocurrent vs time-delay for a preparation and detection pulse both tuned to the bright neutral exciton transition and comparing co and cross-circular excitation. For more details on this technique, see ref. \cite{Khatab_jpcs}. For a gate-voltage of 0.84~V, the fine-structure splitting of the bright-exciton is $\delta_1 = 16.5\pm 0.5~\mathrm{\mu eV}$ and $K=+0.130\pm 0.004 ~\mathrm{\mu eV.T^{-2}}$. The electron tunneling rate is deduced from fits to the total hole population, as presented in fig. \ref{fig:fig4}. Radiative recombination is neglected, since $\Gamma_e \gg \Gamma_r$. The mixing angle of the bright and dark exciton is deduced from single pulse photocurrent spectra data as presented in fig. \ref{fig:fig2}, where the ratio of the dark to bright neutral exciton peaks is assumed to be $\tan^2{(\theta)}\approx (aB)^2$ with $a=0.085\pm 0.010 ~\mathrm{T^{-1}}$.

As can be seen in fig. \ref{fig:fig7}, the model using independently measured fitting parameters gives a reasonable description of both the magnetic and electric-field dependence of the contrast. The model treats the preparation pulse as a delta-function, and for fast electron tunneling times this approximation is less valid. This may explain the departure between experiment and model at high electric-field.

\begin{figure}
\begin{center}
\includegraphics[scale=2.4]{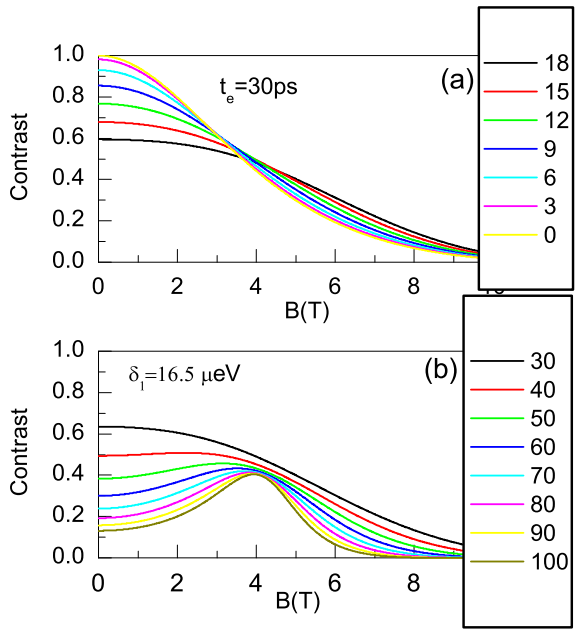}
\end{center}
\caption{Calculations of contrast based on model as a function of (a) fine-structure splitting $\delta_1~\mathrm{(\mu eV)}$ and (b) electron tunneling  time $t_e ~\mathrm{(ps)}$.
}\label{fig:fig8}
\end{figure}

To explore the potential performance of the spin preparation technique, the contrast is calculated for a hypothetical dot and is presented in fig. \ref{fig:fig8}. In fig. \ref{fig:fig8}(a) the dependence on the fine-structure splitting $\delta_1$ is calculated. High contrast spin preparation could be achieved using a dot with zero fine-structure splitting at low magnetic field. For higher fine-structure splitting, the contrast curve exhibits a resonance when the effective fine-structure and hole Zeeman energies are matched.

\section{Conclusions}

The Larmor precession of a single quantum dot hole-spin in a Voigt geometry magnetic field is investigated as a function of electric and magnetic fields. To prepare the hole spin, first a bright neutral exciton is prepared using resonant excitation with circularly polarized light. When the electron tunnels from the dot it leaves a spin-polarized hole that precesses in the Voigt geometry magnetic field. According to a model that describes the contrast of the Larmor precession, the spin preparation is limited by the dynamics of the exciton spin. In the presence of the electron, the hole acquires both a rotating and counter-rotating phase-factor. When the effective fine-structure splitting of the bright neutral exciton matches the hole Zeeman energy, the rotating term of the exciton spin is synchronized with the Larmor precession of the hole spin, whereas for the counter-rotating term the hole-spin is completely dephased leading to a maximum contrast of one half. The spin preparation mechanism bears some similarities to the Hanle effect. In principle, higher contrasts can be achieved at low magnetic fields with zero fine-structure dots. Dots with zero fine-structure can be achieved through annealing \cite{Young_prb}, or applying an in-plane magnetic field \cite{Stevenson_prb}. Recently, Boyer de la Giroday {\it et~al} \cite{Boyer_prl} have proposed a quantum optical memory device based on the electrical separation of a spin-polarized neutral exciton in a double quantum dot device. This work lends some support to the notion that the coherence of the spin state could survive a tunneling event.

\section{Acknowledgements}

We thank the EPSRC (UK) EP/G001642, and the QIPIRC UK  for financial support; H.~Y.~Liu and M.~ Hopkinson for sample growth; and an insightful comment from A.~I.~Tartakovskii regarding the voltage tuneable g-factor. JP thanks CONACYT-Mexico for a Doctoral Scholarship.

\section{Appendix 1: Experimental setup}

The sample is placed inside a helium-bath cryostat, inside the bore of a 5-T magnet at 4.2~K. The sample is mounted on a low temperature xyz nanopositioner (Attocube) using a L-shaped Aluminum bracket with the sample plane parallel to the magnetic field. To excite the sample, the free-space laser beam is folded using a gold-mirror on a $45^{\circ}$ N-BK7 prism, and then focused using a 4.51-mm focal length plano-convex lens. The gold-mirror had no noticeable effect on the polarization.

The sequence of laser pulses is derived from a single 100-fs Ti:sapphire laser with a 76-MHz repetition rate. The beam is split in three using polarizing beamsplitter cubes in conjunction with a half-waveplate to control the power ratios. In each arm, there is a pulse-shaper, essentially a f=50 cm double spectrometer operated in subtraction mode \cite{Boyle_prb,Weiner_rsi}.  After passing through a delay-stage, each arm is coupled into a single-mode fiber. For stable control of the power, there is a fiber U-bench, or free-space break with a polarizer, variable neutral density (ND)filter and a motorized half-waveplate.  After the U-bench, the polarization is pre-compensated for the birefringence of the fiber using a three spool fiber waveplate. The optical-fibers are combined using sequence of non-polarizing 2x2 couplers.  The beam then passes through an optical window of the cryostat and on to the sample.

\end{document}